\documentclass[twocolumn,english,aps,pra,showpacs]{revtex4}
\usepackage[T1]{fontenc}
\usepackage[latin9]{inputenc}
\usepackage{amsmath}
\usepackage{amssymb}
\usepackage{graphicx}
\usepackage{esint}

\makeatletter

\@ifundefined{definecolor}
 {\usepackage{color}}{}
\usepackage[colorlinks=true,linkcolor=blue,urlcolor=blue,citecolor=blue]{hyperref}

\newcommand{\co}[1]{{ #1}}

\makeatother

\usepackage{babel}
\begin{document}

\title{Loop Structure Stability of a Double-Well-Lattice BEC}

\date{\today}

\author{Hoi-Yin Hui,$^{1,2}$ Ryan Barnett,$^{1,2}$ J. V. Porto,$^{1}$
S. Das Sarma$^{1,2}$}

\affiliation{$^{1}$Joint Quantum Institute and $^{2}$Condensed Matter Theory
Center, Department of Physics, University of Maryland, College Park,
Maryland 20742-4111, USA}
\begin{abstract}
In this work, we consider excited many-body mean-field states of bosons
in a double-well optical lattice by investigating stationary Bloch
solutions to the non-linear equations of motion. We show that, for
any positive interaction strength, a loop structure emerges at the
edge of the band structure whose existence is entirely due to interactions.
This can be contrasted to the case of a conventional optical (Bravais)
lattice where a loop appears only above a critical repulsive interaction
strength. Motivated by the possibility of realizing such non-linear
Bloch states experimentally, we analyze the collective excitations
about these non-linear stationary states and thereby establish conditions
for the system's energetic and dynamical stability. We find that there
are regimes that are dynamically stable and thus apt to be realized
experimentally. 
\end{abstract}

\maketitle

\section{Introduction}

A wealth of non-trivial quantum states have been realized with Bose-Einstein
condensates (BECs) in optical lattices \cite{Lewenstein2007Ultracold,Bloch2008Many-body}.
Conventionally, these states occupy the lowest energy configuration
of the parent Hamiltonian of the system. On the other hand, motivated
by the low levels of dissipation in ultracold atomic gases, considerable
recent effort has shifted to the possibility of realizing interesting
quantum states not as the ground state, but as long-lived excited
states \cite{Haller2009Realization}. In particular, recent experiments
have realized condensates in excited bands \cite{Muller2007State,Wirth2010Evidence,Heinze2011Multiband,Olschlager2011Unconventional,Soltan-Panahi2011Quantum}.
Such efforts open the door to achieving, for instance, states having
Neél order, which is notoriously difficult to realize as the ground
state of a system of bosons \cite{Sorensen2010Adiabatic,Hui2011Instabilities},
and novel fermionic states \cite{Sun2011Topological,Li2012Topological}
that have no parallels in solid state physics. Unconventional excited
states realized with ultracold atoms have been considered theoretically
in Refs.~\cite{Wu2006Quantum,Wu2008Theory,Zhao2008Orbital,Umucalilar2008p,Cai2011Complex,Lewenstein2011Optical}.

In the presence of interactions, the nonlinear Bloch band of a BEC
can develop interesting features. Although its ground state is fairly
conventional, at the Brillouin zone boundary, the band can develop
a cusp and subsequently form a loop as interaction is increased \cite{Wu2000Nonlinear,Diakonov2002Loop,Machholm2004Spatial,Seaman2005Period}.
However, for Bravais lattices (with one lattice site per unit cell)
such a loop appears only above a critical interaction strength, which
can be large. For all realistic situations, the loop is also small
and hence difficult to detect. This limits the proposed experimental
detection of the loops to only indirect signals such as the nonlinear
Landau-Zener effect \cite{Wu2000Nonlinear,Liu2002Theory,Choi2003To}.

Similar looped band structures also appear for BEC on a double-well
optical lattice \cite{Seaman2005Period,Witthaut2011Quantum}. Unlike
the conventional looped structure described above, however, we find
that a significantly large loop is induced for any interaction strength,
and a large energy separation from the excited band is possible by
suitably tuning the lattice depth and the lattice staggering. With
the recent experimental realization of double-well optical lattices
\cite{Sebby-Strabley2006Lattice,Lee2007Sublattice,Wirth2010Evidence}
and the concomitant theoretical investigations of the many-body phenomena
in them \cite{Danshita2007Quantum,Peterson2008Realizing,Stojanovic2008Incommensurate,Vaucher2008Creation,Yukalov2008Cold,Wagner2011Spin-1,Wang2011Mott-insulating,Zhou2011Interaction-induced},
it is now an appropriate moment to consider the possibility of detecting
the looped band on a double-well lattice.

Previously, the interaction-induced loop structure in double-well
optical lattice was considered theoretically for some specific cases.
In Ref.~\cite{Seaman2005Period}, a one-dimensional Kronig-Penney
potential was used to demonstrate period-doubling in a double-well
lattice in one dimension. The special form of the potential allowed
analytic solutions to be obtained. A tight-binding model of a one-dimensional
double-well system was analyzed in Ref.~\cite{Witthaut2011Quantum}.
Ref.~\cite{Olschlager2009Kinetic} found an analytical solution at
the band edge for a specific interaction energy by employing the Thomas-Fermi
approximation. \co{However, the computation of Bogoliubov spectrum, and hence the
stability, around the states of a two-dimensional loop structure is lacking in the literature.}

In this work, we compute the interacting Bloch solutions of the Gross-Pitaevskii
equation for two-dimensional double-well lattices using realistic
lattice potentials. This allows us to elucidate the behavior of the
loop structure at the band edge as a function of the lattice potential.
Essential to the experimental realization of the loop states is their
stability. By analyzing the behavior of the collective modes about
the mean-field solutions, we find that a range of states in the excited
band are dynamical stable, which are in experimentally accessible
parameter regimes.

The paper is organized as follows. In Sec.~\ref{sec:Hamiltonain},
we set up the problem and indicate the specific lattice used in our
analysis. We then describe the method of obtaining non-linear Bloch
solutions to the mean field equations of motion, and present the loop
band structure in Sec.~\ref{sec:Sol}. In Sec.~\ref{sec:Stability},
we compute the collective excitations about the non-linear Bloch solutions.
These are used to establish the energetic and dynamical stability
of the system prepared in these states. In Sec.~\ref{sec:TB}, we
present a tight-binding model that captures some, but not all, of
the salient features of the more-accurate continuum results. We conclude
in Sec.~\ref{sec:Conclusions}.

\section{The Double-Well Optical Lattice\label{sec:Hamiltonain}}

\begin{figure}
\begin{centering}
\includegraphics[width=1\columnwidth]{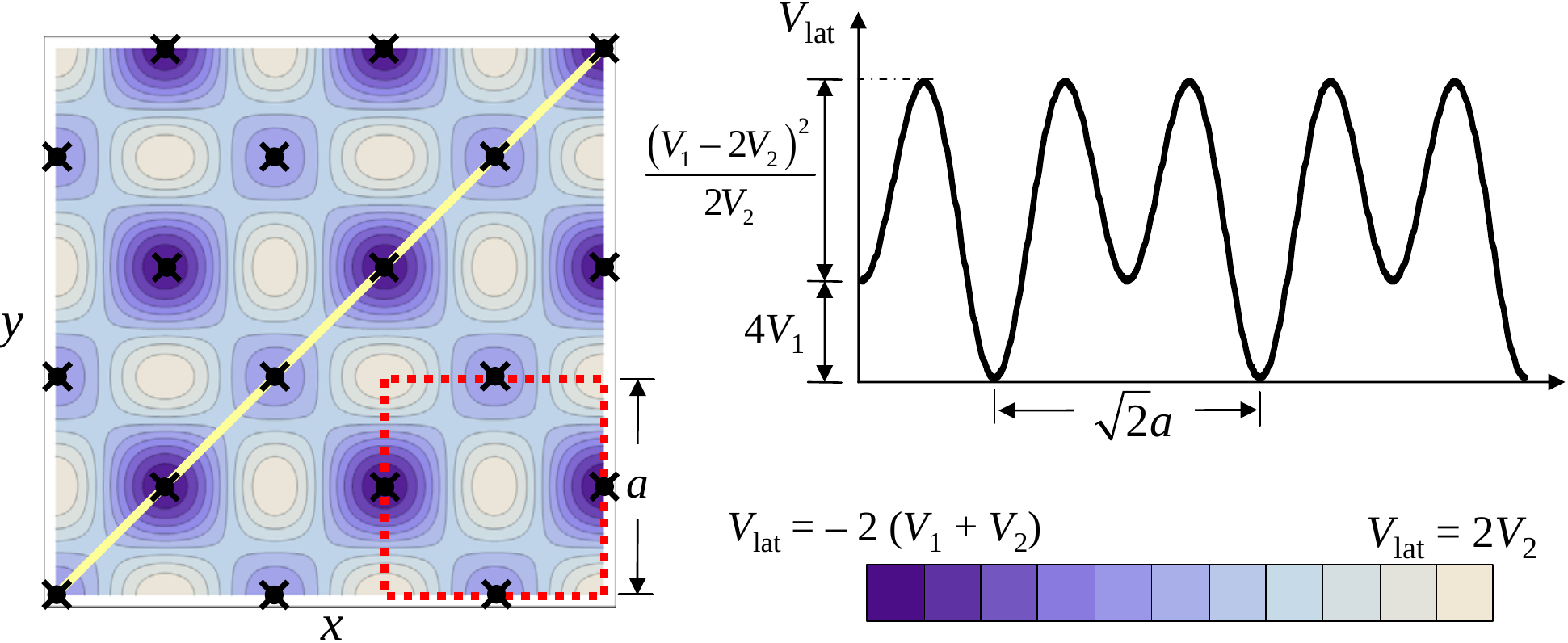} 
\par\end{centering}

\caption{(Left) The double-well lattice potential used in this work. The unit
cell (dotted red square) and lattice constant $a$ are shown. Each
black cross corresponds to a site of a bosonic annihilation operator
in Eq.~(\ref{eq:HTB}). The minima are colored deep-blue (gray),
in accordance to the scale on lower-right. (Upper-right) A slice of
the lattice potential along the diagonal line on the left panel.}

\label{fig:Lattice} 
\end{figure}

In a recent experiment \cite{Sebby-Strabley2006Lattice}, a two-dimensional double-well
lattice was realized by superimposing standing waves in the $x$ and
$y$ directions. To ensure phase stability, a single light source
was used and the lattice is formed in a folded-retroreflected geometry.
We consider a lattice potential that is realizable with this geometry,
given by 
\begin{eqnarray}
V_{\mathrm{lat}}(x,y) & = & V_{1}\left[\cos(2k_{L}x)-\cos(2k_{L}y)\right]\label{eq:lattice}\\
 &  & +2V_{2}\cos(2k_{L}x)\cos(2k_{L}y)\nonumber 
\end{eqnarray}
where $k_{L}=\pi/a$ and $a=\lambda/\sqrt{2}$ is the lattice constant
(shown in Fig.~\ref{fig:Lattice}). By adjusting the path-length differences
of the lattice beams, a more general lattice can be generated. We
further restrict ourselves to the case where $2V_{2}>V_{1}>0$ which
gives double-well lattices with degenerate maxima and staggered local
minima (see Fig.~\ref{fig:Lattice})

With this lattice potential, the full Hamiltonian of the system is
given by 
\begin{eqnarray}
{\cal H} & = & \int d^{3}\mathbf{r}\left\{ \hat{\psi}^{\dagger}(\mathbf{r})\left[\frac{-\hbar^{2}}{2m}\mathbf{\nabla}^{2}+V_{\mathrm{lat}}(\mathbf{r})\right]\hat{\psi}(\mathbf{r})\right.\nonumber \\
 &  & \left.+\frac{g}{2}\hat{\psi}^{\dagger}(\mathbf{r})\hat{\psi}^{\dagger}(\mathbf{r})\hat{\psi}(\mathbf{r})\hat{\psi}(\mathbf{r})\right\} \label{eq:H}
\end{eqnarray}
where $g=\frac{4\pi\hbar^{2}a_{s}}{m}$, $\hat{\psi}(\mathbf{r})$
describes the destruction of a boson at position $\mathbf{r}$, $m$
is the mass of the constituent bosons, and $a_{s}>0$ is the $s$-wave
scattering length.

For the mass and scattering length, we use parameters for $^{87}$Rb.
We consider spatially averaged densities $\bar{\rho}$ below $1\times10^{14}\mbox{cm}^{-3}$,
since for larger densities three-body losses become important. In
the following analysis we restrict our attention to the case where
$V_{1}$ and $V_{2}$ are less than $10E_{R}$, and set the recoil
energy to be $E_{R}=\frac{\hbar^{2}k_{L}^{2}}{2m}=h\times1.75$ kHz,
given by the experimental parameters. For such parameters the tight-binding
approximation is not necessarily valid. For this range of $V_{1}$
and $V_{2}$, the ground state is in the superfluid phase, well away
from the Mott insulating transition.

\section{Mean-field Analysis of the interacting band structure\label{sec:Sol}}

In this Section we describe the numerical method used to obtain periodic
mean-field stationary states of Eq.~(\ref{eq:H}). These solutions
are shown to exhibit extra {}``looped'' states which are then further
analyzed. We concentrate on the semi-classical regime of $\mathcal{H}$
and approximate $\hat{\psi}$ with $\psi\equiv\langle\hat{\psi}\rangle$.
Thus, we seek solutions of the Gross-Pitaevskii equation (GPE) 
\begin{equation}
\mu\psi(\mathbf{r})=\frac{-\hbar^{2}}{2m}\mathbf{\nabla}^{2}\psi(\mathbf{r})+V_{\mathrm{lat}}(\mathbf{r})\psi(\mathbf{r})+g\left|\psi(\mathbf{r})\right|^{2}\psi(\mathbf{r})\label{eq:GPE}
\end{equation}
\co{Here we consider only a 2D system. Qualitatively similar structure should be
expected in other dimensions, but we shall focus on 2D for experimental relevance.}
Among the solutions of this nonlinear differential equation, we are
interested in the solutions of the Bloch form: 
\begin{equation}
\psi_{n\mathbf{k}}(\mathbf{r})=e^{i\mathbf{k}\cdot\mathbf{r}}u_{n\mathbf{k}}(\mathbf{r})
\end{equation}
where $n$ is the band index, $\mathbf{k}$ is the crystal momentum,
and $u_{n\mathbf{k}}({\bf r})$ has the periodicity of the underlying
lattice. The corresponding mean-field energy per unit cell is then

\begin{eqnarray}
E_{n\mathbf{k}} & = & \int_{\mathrm{cell}}d\mathbf{r}\left(-\frac{\hbar^{2}}{2m}\psi_{n\mathbf{k}}^{*}\mathbf{\nabla}^{2}\psi_{n\mathbf{k}}+V_{\mathrm{lat}}(\mathbf{r})\left|\psi_{n\mathbf{k}}\right|^{2}\right.\nonumber \\
 &  & \left.+\frac{g}{2}\left|\psi_{n\mathbf{k}}\right|^{4}\right)
\end{eqnarray}
Throughout the paper, we choose the zero of energy such that the ground
state energy is zero.

\begin{figure}
\begin{centering}
\includegraphics[width=0.7\columnwidth]{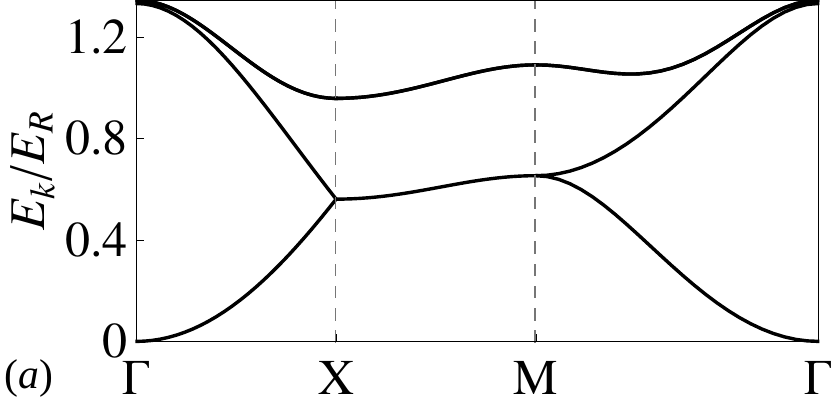} 
\par\end{centering}

\begin{centering}
\includegraphics[width=0.7\columnwidth]{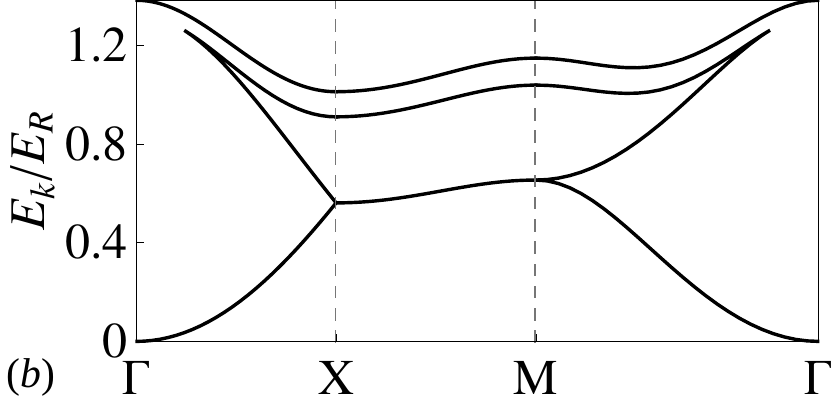} 
\par\end{centering}

\begin{centering}
\includegraphics[width=0.7\columnwidth]{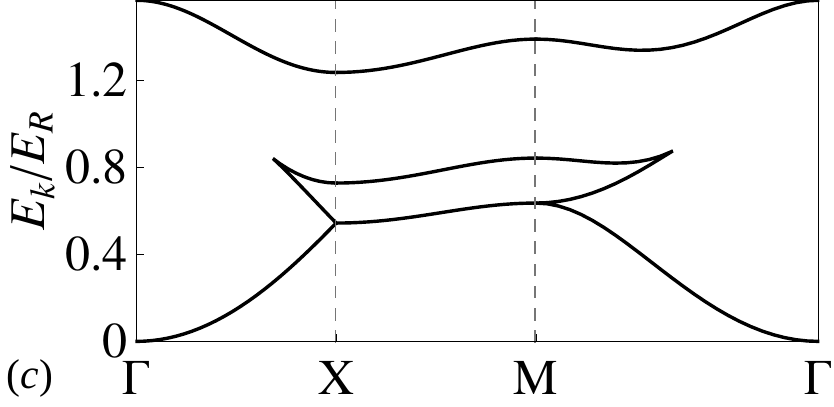} 
\par\end{centering}

\caption{The lowest two bands computed with: $g\bar{\rho}=0.4E_{R}$ and $V_{2}=4.8E_{R}$,
while $V_{1}$ are: (a) $0$ (i.e. no staggering), (b) $0.04E_{R}$,
and (c) $0.2E_{R}$. The labeling of the crystal momenta $\mathbf{k}$
follows the convention where $\Gamma=\left(0,0\right)$, $X=\left(\pi/a,0\right)$
and $M=\left(\pi/a,\pi/a\right)$.}

\label{fig:Band2D} 
\end{figure}

We numerically compute the nonlinear Bloch band by expanding $u_{n\mathbf{k}}(\mathbf{x})$
as $u_{n\mathbf{k}}(\mathbf{x})=\sum_{\mathbf{K}}c_{n\mathbf{K}}e^{i\mathbf{K}\cdot\mathbf{r}}$
where $\mathbf{K}$ are reciprocal lattice vectors and the summation
includes a sufficient number of harmonics to ensure accuracy. This
is then substituted to Eq.~(\ref{eq:GPE}) and a set of coupled equations
is obtained by equating coefficients of equal harmonics. Together
with the normalization condition, $\left\{ c_{n\mathbf{K}}\right\} $
and $\mu$ are then solved-for numerically. We start with an initial
solution $\left\{ c_{n\mathbf{K}}\right\} $ at $\mathbf{k}=\mathbf{0}$,
found with imaginary-time propagation. Then $\mathbf{k}$ is changed
stepwise and $\left\{ c_{n\mathbf{K}}\right\} $ is computed through
numerical root-finding at each step. The non-linear band structures
for several sets of parameters are shown in Fig.~\ref{fig:Band2D}.

The most prominent feature is the emergence of a looped band structure
when a lattice staggering $V_{1}\ne0$ is introduced. Similar phenomena
were also discovered in interacting BECs on regular 1D \cite{Wu2000Nonlinear,Diakonov2002Loop,Machholm2003Band,Machholm2004Spatial,Seaman2005Nonlinear}
and 2D \cite{Chien2010Two-stage,Chen2011Bose-Einstein} lattices.
However, a substantial looped band on a regular (unstaggered) lattice
requires the interaction energy to be much larger than the lattice
depth, which is experimentally a stringent condition. In contrast,
a substantial looped band on a staggered lattice only requires the
interaction energy to be greater than the staggering, which is achievable
as a staggering of $<1\%$ can be realized.

For a qualitative understanding of the emergence of the looped band
on a double-well lattice, we consider the band structures for different
staggering in Fig.~(\ref{fig:Band2D}). First consider the case of
$V_{1}=0$ (Fig.~\ref{fig:Band2D}a), in which case our unit cell
is twice the lattice's natural unit cell. It is known that an extra
solution exists at the band edge \cite{Seaman2005Period} and a staggering
splits the upper and lower band and hence a loop is induced (Fig.~\ref{fig:Band2D}b).
Further increasing the staggering enlarges the energy gap between
the excited band and the loop (Fig.~\ref{fig:Band2D}c). The importance
of using a double-well lattice is apparent in Fig.~\ref{fig:Band2D}c,
where we see a substantial loop well separated from the excited band.

\co{Note that sharp edges are formed in the lowest band, which can lead to 
interesting nonlinear Landau-Zener effects. This was proposed to 
detect the loop band structure experimentally in 1D system \cite{Wu2000Nonlinear,Liu2002Theory,Choi2003To}.
In our current system, we expect similar a effect to appear as one traverses the band edge
from the $\Gamma$ point to the $X$ point.}

\begin{figure}
\begin{centering}
\includegraphics[width=0.9\columnwidth]{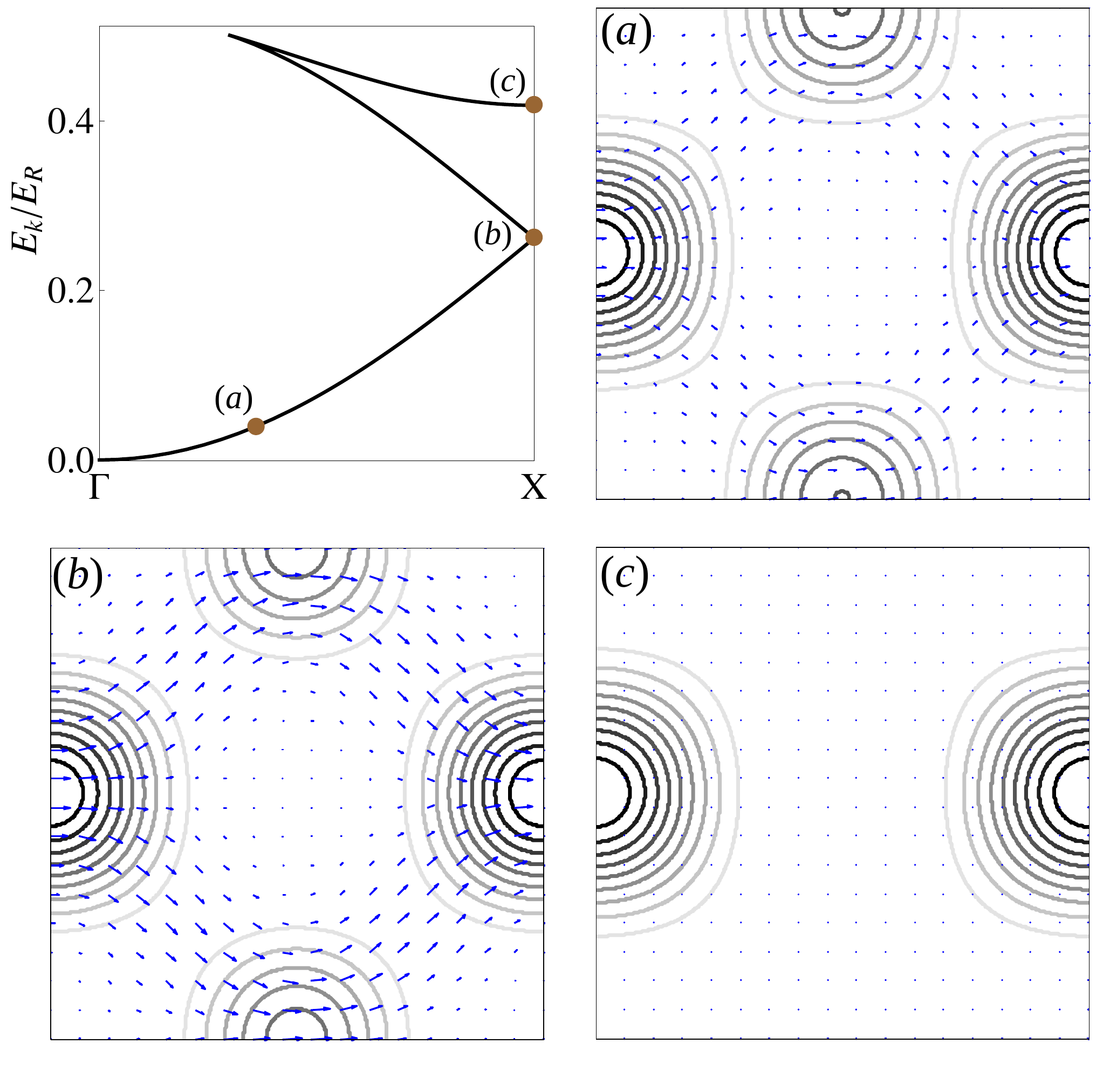} 
\par\end{centering}

\caption{The wavefunction and current flow in a unit cell (shown in Fig.~\ref{fig:Lattice})
for three eigenstates where (a) $\mathbf{k}a=\left(0.18\pi,0\right)$,
(b,c) $\mathbf{k}a=\left(\pi,0\right)$. The corresponding points
on the energy band are shown on the upper-left panel. {[}The other
degenerate solution at point (b) of the band is given by the horizontal
reflection of Fig.~b{]} The other panels show the contours of  $|\psi(\mathbf{x})|$
and the field of $\mathbf{j}$. The parameters are: $g\bar{\rho}=0.4E_{R}$,
$V_{1}=0.4E_{R}$ and $V_{2}=8E_{R}$. \label{fig:flow}}
\end{figure}

The superfluid current density is obtained from the Bloch solution
$\psi_{n\mathbf{k}}(\mathbf{x})$ as 
\begin{align}
\mathbf{j}_{n\mathbf{k}} & =\frac{\hbar}{m}\mathrm{Im}\left(\psi_{n\mathbf{k}}^{*}\mathbf{\nabla}\psi_{n\mathbf{k}}\right)\nonumber \\
 & =\frac{\hbar}{m}\sum_{\mathbf{K}\mathbf{K'}}\left(\mathbf{K}+\mathbf{k}\right)c_{n\mathbf{K'}}^{*}c_{n\mathbf{K}}\cos\left(\mathbf{K'}\cdot\mathbf{r}-\mathbf{K}\cdot\mathbf{r}\right)\label{eq:current}
\end{align}
In Fig.~\ref{fig:flow} we plot the wavefunction and current density
of several indicated states. A currentless state is found at the band
edge of the interaction-induced state (Fig.~\ref{fig:flow}c). For
a non-interacting system, it is well known that the cell-averaged
current satisfies $\bar{\mathbf{j}}_{n\mathbf{k}}=\frac{1}{\hbar}\nabla_{\mathbf{k}}E_{n\mathbf{k}}$.
This relation, in fact, can also be shown to hold for the nonlinear
system in Eq.~(\ref{eq:GPE}). Therefore currentless solutions can
only appear at the $\Gamma$ point or the $(c)$ point in Fig.~(\ref{fig:flow}),
where the energy band attains its local extrema. The state at $(c)$,
however, has identically zero current everywhere instead of just an
average zero current. This is because the period doubled state can
be taken to be real everywhere, and hence currentless by virtue of
Eq.~(\ref{eq:current}). It is easy to see that a real solution is
possible only at the $\Gamma$, $X$ or $M$ points of the Brillouin
zone. Because $\psi(\mathbf{r})=\sum_{\mathbf{K}}c_{\mathbf{K}}e^{i\left(\mathbf{K}+\mathbf{k}\right)\cdot\mathbf{r}}$,
the reality of $\psi$ requires $2\mathbf{k}$ be equal to a reciprocal
lattice vector and that $c_{\mathbf{K}}=c_{-\mathbf{K}-2\mathbf{k}}$.
This is satisfied by the states of the loop at the $X$ and $M$ points.

\section{Stability\label{sec:Stability}}

A crucial factor determining the realizability of the states on the
loop is their stability. In general, there are two qualitatively different
types of instabilities that could potentially occur for interacting
BEC: energetic instability and dynamical instability. Energetic instability
occurs if the system is not at a local minimum of the mean-field energy.
This, however, is often unimportant in experiments since the timescales
for energy dissipation are long. In contrast, when the system has
collective fluctuations with complex frequencies, small perturbations
will grow exponentially fast: a dynamical instability. This type of
instability does not require energy dissipation, and it will cause
rapid depletion and fragmentation of the condensate \cite{Fallani2004Observation,Ferris2008Dynamical}.
Thus realizing a dynamically unstable state is difficult if the instability
timescale is much shorter than the experimental timescale.

\begin{figure}
\begin{centering}
\includegraphics[width=0.8\columnwidth]{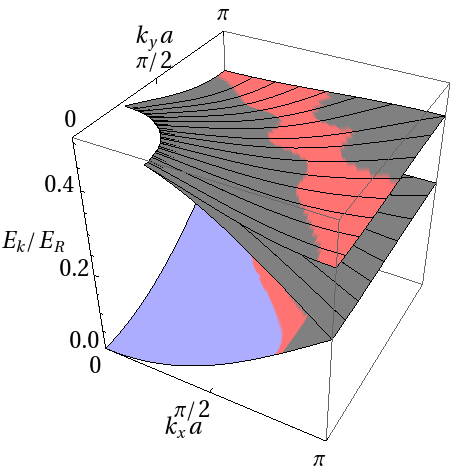} 
\par\end{centering}

\caption{(Color online) Nonlinear Bloch bands with the same parameters of Fig.~(\ref{fig:flow}),
in the first quadrant of the Brillouin zone. States in blue region
are stable while states in gray regions are unstable, both dynamically
and energetically. Red regions are energetically unstable but dynamically
stable. The excited band is separated from the ground band by about
$1E_{R}$ and is not shown here. \label{fig:Band3D}}
\end{figure}

\co{To analyze the stability of states \cite{Wu2000Nonlinear}, it is therefore necessary
to consider the fluctuation modes about the metastable excited states (every point on the 
nonlinear Bloch band). We follow the standard procedure for computing the
Bogoliubov spectrum. That is,} we expand the Hamiltonian to quadratic
order in the field: $\hat{\psi}\rightarrow\psi_{n\mathbf{k}}+e^{i\mathbf{k}\cdot\mathbf{r}}\hat{\phi}(\mathbf{x})$
where $\psi_{n\mathbf{k}}$ is the stationary non-linear Bloch solution.
Hence the term linear in $\hat{\phi}$ vanishes and the term of second
order in $\hat{\phi}$ is: 
\begin{eqnarray}
\delta^{2}\mathcal{H} & = & \sum_{\mathbf{K}\mathbf{q}}\left[\frac{\left(\mathbf{k}+\mathbf{K}+\mathbf{q}\right)^{2}}{2m}\hat{\phi}_{\mathbf{q}+\mathbf{K}}^{\dagger}\hat{\phi}_{\mathbf{q}+\mathbf{K}}+V_{\mathbf{K}-\mathbf{K}'}\hat{\phi}_{\mathbf{q}+\mathbf{K}}^{\dagger}\hat{\phi}_{\mathbf{q}+\mathbf{K}'}\right.\nonumber \\
 &  & \left.+2g\rho_{\mathbf{K}-\mathbf{K}'}\hat{\phi}_{\mathbf{q}+\mathbf{K}}^{\dagger}\hat{\phi}_{\mathbf{q}+\mathbf{K}'}\right.\label{eq:fluct}\\
 &  & \left.+g\left(\tilde{\rho}_{\mathbf{K}-\mathbf{K}'}^{*}\hat{\phi}_{\mathbf{q}+\mathbf{K}}\hat{\phi}_{-\mathbf{q}-\mathbf{K}'}+\tilde{\rho}_{\mathbf{K}-\mathbf{K}'}\hat{\phi}_{\mathbf{q}+\mathbf{K}}^{\dagger}\hat{\phi}_{-\mathbf{q}-\mathbf{K}'}^{\dagger}\right)\right]\nonumber 
\end{eqnarray}
 where $\hat{\phi}_{\mathbf{q}}$ is the Fourier transform of $\hat{\phi}(\mathbf{x})$.
$\rho=\left|\psi_{n\mathbf{k}}\right|^{2}$ is an effective potential
created by the BEC, and $\tilde{\rho}=\psi_{n\mathbf{k}}^{2}$. The
\co{Bogoliubov spectrum} is then numerically found by canonical transformation
of Eq.~(\ref{eq:fluct}). For each crystal momentum $\mathbf{k}$,
we compute the spectrum $\omega_{\mathbf{q}}$. The presence of any
complex $\omega_{\mathbf{q}}$ implies both dynamical and energetic
instabilities, while negative $\omega_{\mathbf{q}}$ indicates an
energetic instability. The stabilities of the Bloch states are shown
in Fig.~\ref{fig:Band3D}. We find the top of the loop has a region
of dynamically stable states, which trace out a roughly circular path
in the Brillouin zone.

\begin{figure}
\begin{centering}
\includegraphics[width=0.85\columnwidth]{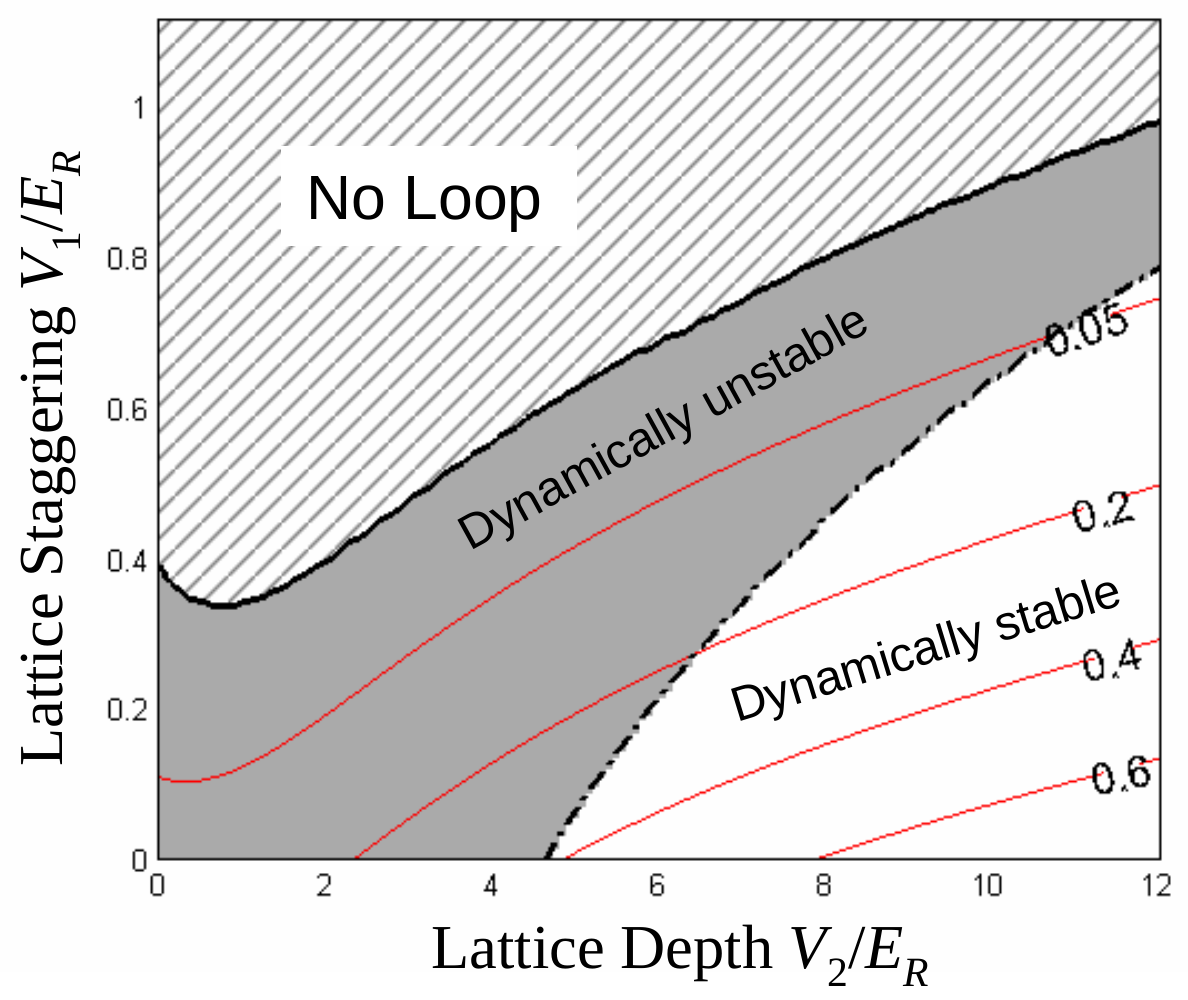} 
\par\end{centering}

\caption{Stability of the eigenstate at $\mathbf{k}a=\left(\pi,0\right)$ on
the loop, as a function of $V_{1}$ (the staggering) and $V_{2}$
(the lattice depth) in Eq.~(\ref{eq:lattice}), with $g\bar{\rho}=0.4E_{R}$
(corresponding to a density $\bar{\rho}=9\times10^{13}{\rm cm}^{-3}$).
The hashed region has no loop. The shaded region is dynamically unstable
and the white region is energetically unstable but dynamically stable.
Contours show the loop size (i.e. vertical distance between (b) and
(c) in Fig.~\ref{fig:flow}) in units of $E_{R}$. \label{fig:Phase}}
\end{figure}

The stability phase diagram of the state at the tail at $\mathbf{k}a=(\pi,0)$
as a function of $V_{1}$ and $V_{2}$ is shown in Fig.~\ref{fig:Phase}.
It is divided into three regions: (1) when the lattice staggering
is sufficiently large, the loop is totally suppressed; (2) as the
staggering becomes smaller, a loop is formed but with all states on
it dynamically unstable; (3) further reducing staggering enlarges
the loop and a band of stable states is formed on it, of which Fig.~\ref{fig:Band3D}
is an example.

\section{Tight-Binding Model\label{sec:TB}}

In this Section, we use a tight-binding model to understand some of
our previous results, in the limit of large lattice potentials. The
tight-binding Hamiltonian we consider is 
\begin{eqnarray}
\mathcal{H}_{\mathrm{TB}} & = & -t\sum_{\left\langle \mathbf{r}\mathbf{r}'\right\rangle }\left(\hat{b}_{\mathbf{r}}^{\dagger}\hat{b}_{\mathbf{r}'}+\mbox{h.c.}\right)+\Delta\sum_{\mathbf{r}}e^{i\mathbf{Q}\cdot\mathbf{r}}\hat{b}_{\mathbf{r}}^{\dagger}\hat{b}_{\mathbf{r}}\nonumber \\
 &  & +\frac{U}{2}\sum_{\mathbf{r}}\hat{b}_{\mathbf{r}}^{\dagger}\hat{b}_{\mathbf{r}}^{\dagger}\hat{b}_{\mathbf{r}}\hat{b}_{\mathbf{r}}\label{eq:HTB}
\end{eqnarray}
where $\hat{b}_{\mathbf{r}}$ annihilates a boson at position $\mathbf{r}$
on a square lattice with unit lattice constant, $\sum_{\left\langle \mathbf{r}\mathbf{r}'\right\rangle }$
indicates summation over nearest neighbors, and $\mathbf{Q}=\left(\pi,\pi\right)$.
The hopping amplitude is given by $t$, $\Delta>0$ is the on-site
energy staggering between sites, while $U$ denotes the on-site interaction
energy between bosons. This model describes Eq.~(\ref{eq:H}) in
the limit of a strong lattice potential, where we associate a bosonic
annihilation operator to each local minimum of the lattice (see Fig.~\ref{fig:Lattice}).
Note that a term of staggered hopping is unnecessary because the lattice
has 4-fold rotational symmetry and all nearest-neighbor links are
equivalent.

In the superfluid regime, the system is described by a mean-field
equation $\mu b_{\mathbf{r}}=-t\sum_{\mathbf{\delta}}b_{\mathbf{r}+\mathbf{\delta}}+\Delta e^{i\mathbf{Q}\cdot\mathbf{r}}b_{\mathbf{r}}+U\left|b_{\mathbf{r}}\right|^{2}b_{\mathbf{r}}$,
where $b_{\mathbf{r}}=\langle\hat{b}_{\mathbf{r}}\rangle$ and $\mbox{\ensuremath{\delta}}$
denote the positions of the nearest-neighbors. This is solved by
\begin{equation}
b_{\mathbf{r}}=\sqrt{\rho}\left(\cos\frac{\alpha_{\mathbf{k}}}{2}+e^{i\mathbf{Q}\cdot\mathbf{r}}\sin\frac{\alpha_{\mathbf{k}}}{2}\right)e^{i\mathbf{k}\cdot\mathbf{r}}
\end{equation}
where $\alpha_{\mathbf{k}}$ satisfies
\begin{equation}
2t\left(\cos k_{x}+\cos k_{y}\right)\sin\alpha_{\mathbf{k}}+\Delta\cos\alpha+U\rho\sin\alpha_{\mathbf{k}}\cos\alpha_{\mathbf{k}}=0.\label{eq:alphaEq}
\end{equation}
This yields the nonlinear Bloch band $E_{\mathbf{k}}/\rho=\Delta\sin\alpha_{\mathbf{k}}-2t\left(\cos k_{x}+\cos k_{y}\right)\cos\alpha_{\mathbf{k}}-U\rho\left(\frac{1}{2}\cos^{2}\alpha_{\mathbf{k}}-1\right)$.

In vanishing interaction $(U=0)$, this tight-binding model generates
two bands. When $U$ is raised above a critical threshold, however,
four solutions appear in some regions of $\mathbf{k}$ near the band
edge. These are the extra states of the loop. Since Eq.~(\ref{eq:alphaEq})
has four solutions if $\left(U\rho\right)^{2/3}>\Delta^{2/3}+\left(2t\left(\cos k_{x}+\cos k_{y}\right)\right)^{2/3}$,
the condition to have a loop without it filling up the entire Brillouin
zone (hence forming a separate band) is therefore $\Delta<U\rho<\left(\Delta^{2/3}+\left(4t\right)^{2/3}\right)^{3/2}$.
Further, by demanding all Bogoliubov modes to be real, we find that
the condition of having the state at $\mathbf{k}=\left(\pi,0\right)$
on the loop dynamically stable is $t<\frac{\left(Un-\Delta\right)^{3/2}}{8\sqrt{Un+\Delta}}$.
Since in our lattice {[}Eq.~(\ref{eq:lattice}){]} $V_{1}$ controls
the amount of staggering and $V_{2}$ controls lattice depth, the
phase diagram in the continuum (Fig.~\ref{fig:Phase}) is in qualitative
agreement with the tight-binding model. The wavefunction at the band
edge $\left(\cos k_{x}+\cos k_{y}=0\right)$ can be analytically solved.
When $U\rho>\Delta$, the state on the loop has $\alpha=-\frac{\pi}{2}$.
Thus the state at the tail is $b_{\mathbf{r}}=\sqrt{2\rho}$ if $e^{i\mathbf{Q}\cdot\mathbf{r}}=-1$
and zero otherwise, i.e. has all particles on the lower wells. This
too agrees with the continuum calculations in Fig.~\ref{fig:flow}.

This loop structure could also be understood from the period-doubled
solution \cite{Seaman2005Period}. It is known that this tight-binding
model without staggering $\left(\Delta=0\right)$ admits period-$p$
solutions, where $p$ is any positive integer. In particular, a period-doubled
solution $\frac{E_{\mathbf{k}}}{\rho}=\frac{4t^{2}\left(\cos k_{x}+\cos k_{y}\right)^{2}}{2U\rho}+U\rho$
always exists near the band edge \cite{Machholm2004Spatial}. If we
introduce small staggering $\Delta$, the extra band splits in a manner
that generates a looped band structure (Fig.~\ref{fig:Band2D}).
As the staggering gets larger, the splitting between the two bands
also increases but simultaneously suppressing the loop size. As splitting
becomes large $\Delta>U\rho$, the whole loop structure is destroyed.

The adequacy of tight-binding models in describing our loop structure
could be compared with the situations in purely interaction-induced
loop structure \cite{Diakonov2002Loop,Chen2011Bose-Einstein}. In
those systems, a tight-binding model can never produce the loop structure
in continuum regardless of the lattice depth, and this was attributed
to an inappropriate choice of Wannier functions \cite{Chen2011Bose-Einstein}.
In our system, although a loop structure can be captured with the
tight-binding model, some qualitative features of the continuum calculations
are missing in the tight-binding model. For instance, the tight-binding
model predicts that a dynamically stable region on the loop, if present,
must include the band edge. This is not consistent our continuum calculations
in Sec.~\ref{sec:Sol}, as shown in Fig.~\ref{fig:Band3D}.

Although the tight-binding model is simple and gives a physical picture,
the continuum calculations are more relevant from an experimental
viewpoint. This is because the experimentally tunable parameters are
$V_{1}$, $V_{2}$ and $\bar{\rho}$, and it is difficult to compute
from these the suitable tight-binding parameters $t$, $\Delta$ and
$U$ of Eq.~(\ref{eq:HTB}) as shown in \cite{Modugno2011Maximally}.
In contrast, the results we obtained for continuum Hamiltonian Eq.~(\ref{eq:H})
can be directly compared with experimental results. There are also
experimentally-relevant regimes considered above, where a single-orbital
tight-binding approximation is not correct.

\section{Conclusions\label{sec:Conclusions}}

In this work, we have numerically computed the nonlinear Bloch band
structure for an interacting BEC on a 2D double-well optical lattice. 
We also computed the Bogoliubov modes about all the states, and thereby
mapped out the stability phase diagram as a function of lattice depth
and staggering for experimentally realistic parameters. A tight-binding
model was also considered and it is found to reproduce some qualitative
features of the system.

We find that the interaction energy required to create a
looped band is smaller on a double-well lattice, compared
with a Bravais lattice. Further, we find a stable state on
the loop can be realized within an experimental accessible
parameter range. \co{This raises the possibility of directly
exciting a BEC onto the loop with Raman excitation and/or dynamic manipulation
of the lattice structure \cite{Sebby-Strabley2006Lattice,Wirth2010Evidence}.
Time-of-flight spectroscopy of the momentum distribution
could be used to experimentally confirm the unique loop structure.}

In future work, a more careful treatment of the important possibility
of Mott correlations about the states on the loop will be interesting
to study. When the system is in the tight-binding regime, the Gutzwiller
method (which is more general than the treatment provided in Sec.~\ref{sec:TB})
will be suitable for this purpose. The effects of the non-linear Bloch
states on the dynamics of BECs in double-well lattices will also be
of interest.

\acknowledgments

We thank I.B. Spielman and S. Powell for critically reading the manuscript.
This work was supported by JQI-NSF-PFC, AFOSR JQI-MURI, ARO Atomtronics
MURI and the DARPA OLE program.

\vfill{}

\bibliographystyle{apsrev4-1}
\bibliography{dw}

\end{document}